\documentclass[hyper]{JHEP} 

\usepackage{epsfig}

\newcommand\fverb{\setbox\pippobox=\hbox\bgroup\verb}

\newcommand\fverbdo{\egroup\medskip\noindent%

            \fbox{\unhbox\pippobox}\ }

\newcommand\fverbit{\egroup\item[\fbox{\unhbox\pippobox}]}

\newbox\pippobox

\title{D-brane Description of New Open String Solutions
in $AdS_5$}
\author{by J. Kluso\v{n}\\
     Department of Theoretical Physics and Astrophysics\\
                   Faculty of Science, Masaryk University\\
Kotl\'{a}\v{r}sk\'{a} 2, 611 37, Brno\\
Czech Republic\\
    E-mail: \email{klu@physics.muni.cz}}
\preprint{\hepth{0805.4719}}

 \abstract{In this paper we find
 D-brane descriptions of some of new
open string solutions that were found
in 0804.3438[hep-th]. These D5-brane
and D3-brane configurations give
gravitational dual descriptions of
Wilson loops in some particular
representations.}

 \keywords{D-branes, AdS/CFT correspondence}

\def\mF{\mathcal{F}}

\def\bA{\mathbf{A}}

\def\bAi{\left(\mathbf{A}^{-1}\right)}

\begin{document}
\section{Introduction and Summary}\label{first}
Wilson loop operators are  non-local gauge invariant operators in
gauge theory in which the theory can be formulated.
 Mathematically we define  a Wilson loop as
the trace in an arbitrary
representation $R$ of the gauge group
$G$ of the holonomy matrix associated
with parallel transport along a closed
curve $C$ in spacetime. Further, as it
is well known from the times of birth
of AdS/CFT correspondence
\cite{Maldacena:1997re} that Wilson
loops in $N=4$ SYM theory can be
calculated in dual description using
macroscopic strings
\cite{Maldacena:1998im,Rey:1998ik}.
This prescription is based on a picture
of the fundamental string ending on the
boundary of
 AdS along the path $C$ specified by the Wilson loop
operator. The description of the Wilson loop in terms
 of a fundamental string is a well established
part of the AdS/CFT dictionary.
Remarkably it was argued in very
interesting paper \cite{Gomis:2006sb}
that Wilson loops have a gravitational
dual description in terms of D5-branes
or alternatively in terms of D3-branes
in $AdS_5\times S^5$ \footnote{For
related works, see
\cite{Bonelli:2008rv,Chen:2008ds,Drukker:2007qr,Lunin:2007zz,Sakaguchi:2007ea,
Chu:2007pb,Chen:2007ir,Gomis:2006mv,Drukker:2006zk,Gomis:2006im,
Armoni:2006ux,Hartnoll:2006ib,Drukker:2006ga,Lunin:2006xr,Hartnoll:2006hr}.}.
More precisely, Drukker and Fiol
\cite{Drukker:2005kx}  argued that a
Wilson loop with matter in the rank $k$
\emph{symmetric representation} is
better described as a D3-brane embedded
in $AdS_5$ with $k$ units of electric
flux. Further, it was argued in
\cite{Hartnoll:2006hr,Hartnoll:2006ib,Yamaguchi:2006tq}
 that a Wilson loop with
matter in the rank $k$
\emph{antisymmetric representation} is
better described by a D5-brane whose
world-volume is a minimal surface in
the AdS part of the geometry times an
$S^4$ inside the $S^5$ and that
supports $k$-units of world-volume
electric  flux.

In recent interesting paper
\cite{Ishizeki:2008dw} new open string
solutions in $AdS_5$ were found.
Further, it was shown that these string
solutions end at the boundary on
various Wilson lines. Since these
solutions correspond to the solutions
of the fundamental string equations of
motion these configurations describe
Wilson loops in fundamental
representations. Then it is natural to
ask the question whether we can find
similar solutions for D3 and D5-brane
that should correspond to the Wilson
loops in antisymmetric and symmetric
representations. We will argue that it
is really possible to find such
configurations.

We begin with the study of D5-brane
that wraps $S^4$ in $S^5$ and that
carries electric flux $\Pi$ along its
embedding into $AdS_5$ . Following
\cite{Hartnoll:2006ib} we consider
D5-brane with Euclidean world-volume.
Then we shown that the solutions of the
equations of motion correspond to the
solution found in the first part of the
paper \cite{Ishizeki:2008dw}. Further,
when we evaluate the action on this
solution we obtain regularized value of
the action that corresponds to the
Wilson line in level $\Pi$
anti-symmetric representation. This
result is in nice agreement with
exception.

As the second example we study
Euclidean D3-brane that wraps $S^2$
subspace of $S^5$. Even if this
D3-brane configuration does not
correspond to the Wilson line in
symmetric representation it is still
instructive to find the solutions of
equation of motion and evaluate
corresponding on-shell action.

Finally we consider D3-brane that
carries electric flux $\Pi$ and that is
embedded in $AdS_5$ completely. We
study this action in Minkowski
signature in order to include
light-cone world-volume coordinates. We
find D3-brane configuration that gives
gravitation dual description of
 Wilson loop in
$\Pi$ level symmetric representation.
On the other hand we also find that the
on-shell action vanishes corresponding
to the expectation value of the
corresponding Wilson line $<W>=1$. This
is in complete agreement with the
result presented in
\cite{Ishizeki:2008dw}.

Let us outline our paper. We explicitly
found D3 and D5-brane configurations
that are analogue of fundamental string
solutions given in
\cite{Ishizeki:2008dw} and that
correspond to dual Wilson loop in level
$\Pi$ anti-symmetric and symmetric
representations.  We mean that this
agreement further supports the claim
that D3-branes and D5-branes
configurations that end on the boundary
of $AdS_5$ correspond to expectation
values of some non-local operators in
dual $N=4$ SYM theory.

This paper can be extended in several
ways. First of all it would be
certainly interesting to find D3-brane
configurations that correspond to the
generalization of the circular Wilson
line as in \cite{Ishizeki:2008dw}. This
is very interesting problem since
description of circular Wilson loops in
terms of D3-branes is rather
non-trivial as was shown in
\cite{Drukker:2005kx}. Further it would
be certainly interesting to find
D3-brane and D5-brane analogue of the
fundamental string solutions that were
studied in second part of paper
\cite{Ishizeki:2008dw}. We hope to
return to these problems in future.

The organization of this paper is as
follows. In the next section
(\ref{second}) we introduce basic
notation and study D5-brane and
D3-brane configurations that wrap $S^4$
subspace and $S^2$ subspace of $S^5$
respectively. Then in section
(\ref{third}) we study D3-brane
configuration where D3-brane is
embedded in $AdS_5$ only. Finally for
reader's convenience  we list
in appendix
(\ref{appendix})
the form of
equations of motion for Dp-brane in
general background.
\section{D5-brane on $S^4$  and D3-brane on $S^2$}
\label{second}
In this section we find D5-brane and D3-brane
configurations that are related to fundamental
string solutions studied in \cite{Ishizeki:2008dw}.

To begin with we write  the metric
of $AdS_5$ and $S^5$ in the form
\begin{equation}
ds^2=\frac{R^2}{z^2}
(\eta_{\mu\nu}dx^\mu dx^\nu+dz^2) +R^2
(d\theta^2+\sin^2\theta d\Omega^{[4]})
\ ,
\end{equation}
where
\begin{equation}
\eta_{\mu\nu}dx^\mu dx^\nu =2dx^+dx^-
+(dx^1)^2+(dx^2)^2 \ ,
\end{equation}
and where we also introduced two light-cone
variables $x^\pm$ as
\begin{equation}
x^\pm=\frac{1}{\sqrt{2}}(x^3\pm x^0) \ .
\end{equation}
Let us now study dynamics of D5-brane
in given background. Note that
 D5-brane action in Euclidean
signature takes the form
\begin{eqnarray}
S&=&\tau_5\int d^{6}\xi
\sqrt{\det\bA}-i\tau_5 \int
(2\pi\alpha')F\wedge C^{(4)} \ ,
\nonumber \\
\bA_{\alpha\beta}&=&\partial_\alpha X^M
\partial_\beta X^N g_{MN}+(2\pi\alpha')F_{\alpha\beta} \ ,
\nonumber \\
\tau_5 &=& \frac{N\sqrt{\lambda}}{8\pi^4 R^6} \ , \quad
R^4=\lambda \alpha' \ ,
\nonumber \\
\end{eqnarray}
where $N$ are units of flux of the Ramond-Ramond five-form,
and where we introduced
Ramond-Ramond four-form   $C^{(4)}$
\begin{equation}
C^{(4)}=R^4(\frac{3}{2}(\theta-\pi)
-\sin^3\theta\cos\theta-\frac{3}{2}\cos\theta
\sin\theta)\mathrm{Vol} S^4\equiv R^4C(\theta)\mathrm{Vol} S^4 \ .
\end{equation}
Now we  propose following ansatz
\begin{eqnarray}\label{D5}
& & x^+=u(\tau,\sigma) \ , \quad
x^1=\xi^0 \equiv \tau \ , \quad
 z=\xi^1\equiv \sigma \ ,   \nonumber \\
&  & x^2=a \ , \quad \theta=\theta_\Pi, \quad \phi^a=\xi^a \ ,
\quad a=1,2,3,4 \ , \nonumber \\
\end{eqnarray}
where $\phi^a$ label coordinates on
$S^4$ and where $\theta_\Pi$ is a
constant that depends on the electric
flux in $\tau,\sigma$ direction.
It turns out that it is necessary to
consider complex gauge field so that
$(2\pi\alpha')F \rightarrow
i(2\pi\alpha')F$. Then
 the
matrix $\bA$ takes the form
\begin{eqnarray}
\bA_{\tau\tau}&=&\frac{R^2}{\sigma^2} \ , \quad
\bA_{\sigma\sigma}=\frac{R^2}{\sigma^2}
\ , \nonumber \\
\bA_{\tau\sigma}&=& i2\pi\alpha'
F_{\tau\sigma} \ , \quad
\bA_{\sigma\tau}=i2\pi\alpha'
F_{\sigma\tau} \ ,  \nonumber \\
\det \bA_{ij}&=& R^8\sin^8\theta\det
\tilde{g}_{ij} \ ,
\nonumber \\
\det \bA &=& R^{12}(\frac{1}{\sigma^4}-
\frac{(2\pi\alpha')^2}{R^4}F^2)\sin^8\theta
\det \tilde{g}_{ij} \ , \quad  F\equiv F_{\tau\sigma} \ ,
 \nonumber \\
\end{eqnarray}
where $\tilde{g}_{ij}$ is the metric on
$S^4$ sphere. Then the equation for
$x^-$ implies
\begin{eqnarray}\label{eqx-}
\partial_\tau[
\frac{\partial_\tau u}{\sigma^2}
\frac{1}{\sqrt{R^4-\sigma^4(2\pi\alpha')F^2}}]+
\partial_\sigma[
\frac{\partial_\sigma u}{\sigma^2}
\frac{1}{\sqrt{R^4-\sigma^4(2\pi\alpha')F^2}}]=0 \ .
\nonumber \\
\end{eqnarray}
Further, the equation of motion for
$A_\alpha$ takes the form
\begin{eqnarray}
\partial_\beta[\bAi_A^{\alpha\beta}
\sqrt{\det\bA}]+\partial_\beta[\epsilon^{\beta\alpha}
C(\theta)]=0 \ .
\nonumber \\
\end{eqnarray}
This equation implies an
existence of the
electric flux $\Pi$
\begin{eqnarray}
\frac{(2\pi\alpha'F)\sin^4\theta}{\sqrt{
\frac{R^4}{\sigma^4}- (2\pi\alpha'
F)^2}}+C(\theta)= \Pi \nonumber \\
\end{eqnarray}
and consequently
\begin{eqnarray}\label{Pif}
(2\pi\alpha')F&=& \frac{(\Pi-C(\theta))
R^2}{\sigma^2\sqrt{\sin^8\theta+(\Pi-C(\theta))^2}}
\ ,
\nonumber \\
\frac{1}{\sqrt{\frac{R^4}{\sigma^4}
-(2\pi\alpha'F^2)}}&=&\frac{\sigma^2
\sqrt{\sin^8\theta+(\Pi-C(\theta))^2}}{R^2\sin^4\theta}
\ .
 \nonumber \\
\end{eqnarray}
Using these results we obtain that the
equation (\ref{eqx-}) takes the form
\begin{eqnarray}
\partial_\tau^2
u-2\frac{1}{u}\partial_\sigma u+
\partial_\sigma^2 u=0 \ .
\nonumber \\
\end{eqnarray}
This has the same form  as the equation
derived in the first part of the paper
\cite{Ishizeki:2008dw}. Further it is
easy to see that the equation of motion
for $x^+$ is automatically satisfied.
Finally, using (\ref{Pif}) we obtain
that the  equation of motion for
$\theta$ takes the form
\begin{eqnarray}
4 \sin^7\theta\cos\theta-4 \sin^4\theta
(\Pi-C(\theta))=0  \nonumber \\
\end{eqnarray}
and consequently we find that $\theta$
depends on the value of electric flux
$\Pi$ as
\begin{eqnarray}\label{Pitheta}
\Pi=\frac{3}{4}\sin 2\theta_\Pi
-\frac{3}{2}(\theta_\Pi-\pi)
\end{eqnarray}
or alternatively
\begin{equation}
 (2\pi\alpha'
F)=-\cos\theta_{\Pi}\frac{R^2}{\sigma^2}
\ .
\end{equation}
Then the action evaluated on this
solution takes the form \footnote{Using
the fact that the volume of unit-$S^4$
sphere is equal to
\[
\mathrm{Vol}S^4=\frac{8}{3}\pi^2 \].}
\begin{eqnarray}
S&=&\tau_5 R^6\int d\mathrm{Vol}S^{(4)}
d\sigma
d\tau\frac{1}{\sigma^2}[\sin^5\theta
-\cos\theta C(\theta)]= \nonumber \\
&=&\frac{N\sqrt{\lambda}}{3\pi^2} \int
d\tau d\sigma \frac{1}{\sigma^2}
[\sin^5\theta -\cos\theta C(\theta)] \
.
\nonumber \\
\end{eqnarray}
The D5-brane solution  extends  to the
boundary of $AdS_5$ and ends there
along a one-dimensional curve. This
opens up the possibility of adding
boundary terms to the action. These
boundary terms do not change the
equations of motion, so the solution is
still the same, but the value of the
action when evaluated at this solution
will in general depend on the boundary
terms. Careful discussion of these
boundary terms was given in
\cite{Drukker:2005kx}. However for our
purposes it is only sufficient to
demand that the resulting action
together with boundary term is gauge
invariant. In fact, as was shown in
\cite{Drukker:2005kx}
in order to achieve gauge invariance we
have to add to the action the term
\begin{eqnarray}
S_{gauge}=-\tau_5 \int d\tau d\sigma d
\mathrm{Vol}S^{(4)}\Pi (2\pi\alpha') F=
- \frac{ \sqrt{\lambda}N}{3\pi^2 R^2}
\int
 d\tau d\sigma \Pi (2\pi\alpha')F=
 \nonumber \\
 =
\frac{\sqrt{\lambda}N}{3\pi^2} \int
d\tau d\sigma \frac{1}{\sigma^2}
\Pi\cos\theta_\Pi=
\frac{\sqrt{\lambda}N}{3\pi^2} \int
d\tau d\sigma
\frac{1}{\sigma^2}(\sin^3\theta_\Pi
\cos^2\theta_\Pi+C(\theta_\Pi)\cos\theta_\Pi)
 \nonumber \\
\end{eqnarray}
and hence
\begin{eqnarray}\label{OnS5}
S_{bulk}+ S_{gauge}=
\frac{N\sqrt{\lambda}}
{3\pi^2}\sin^3\theta_{\Pi} \int d\tau
d\sigma \frac{1}{\sigma^2}= \frac{2N}
{3\pi}\sin^3\theta_{\Pi} S_{funstring} \ ,
 \nonumber \\
\end{eqnarray}
where $S_{funstring}$ is an on-shell
action that has the same form as in
\cite{Ishizeki:2008dw}.
 In other
words we have again find that the
Wilson loop in the $\Pi$-th
antisymmetric representation is given
in terms of the loop in the fundamental
representation as follows from
(\ref{OnS5}). Further, $\theta$ is
related to the electric flux through
(\ref{Pitheta}).

Let us now turn to the study of
D3-brane solutions that  are related to
the solution found in
\cite{Ishizeki:2008dw} and that are
characterized by the property that  the
embedding of D3-brane   corresponds to
the embedding of string world-sheet in
$AdS_5$ and to the embedding
$S^2\hookrightarrow S^5$. We use the
convention given in
\cite{Hartnoll:2006ib} and  write the
relevant metric in the form
\begin{eqnarray}\label{D3met}
ds^2=\frac{R^2}{z^2}
[2dx^+dx^-+dx^idx^i+d\alpha^2+\sin^2\alpha
d\Omega^{[2]}+
\cos^2\alpha d\Omega^{[2]}] \ .  \nonumber \\
\end{eqnarray}
The Euclidean DBI action for D3-brane
takes the form
\begin{equation}
S=\tau_3\int d^4\xi\sqrt{\det\bA} \ ,
\end{equation}
where $\tau_3=\frac{N}{2\pi^2 R^4}$.
Since we are interested in solution
where D3-brane  has two directions in
$AdS_5$ and wraps $S^2
 \hookrightarrow S^5$
we do not need to consider WZ
term. More precisely, let us consider
generalization of the solution given in
\cite{Ishizeki:2008dw} and  propose
following ansatz
\begin{equation}
x_+=u(\tau,\sigma) \ , \quad x^1=\xi^0
\equiv \tau \ , \quad  z=\xi^1\equiv
\sigma, \quad x^2=a \ , \quad
\phi^a=\xi^a \ , \quad a=2,3 \ ,
\end{equation}
where $\phi^a$ label coordinates on the
first subspace  $S^2$  of $S^4$ given
in (\ref{D3met}). Then, as in D5-brane
case  we consider imaginary
$F_{\tau\sigma}$ so that the matrix
$\bA$ takes the form
\begin{eqnarray}
\bA_{\tau\tau}&=&\frac{R^2}{\sigma^2} \
, \quad
\bA_{\sigma\sigma}=\frac{R^2}{\sigma^2}
\ , \nonumber \\
\bA_{\tau\sigma}&=& i2\pi\alpha'
F_{\tau\sigma} \ , \quad
\bA_{\sigma\tau}=i2\pi\alpha'
F_{\sigma\tau} \ ,  \nonumber \\
\det \bA_{ab}&=& R^4\sin^4\alpha\det
\tilde{g}_{ab} \ ,
\nonumber \\
\det \bA &=& R^{8}(\frac{1}{\sigma^4}-
\frac{(2\pi\alpha')^2}{R^4}F^2)\sin^4\alpha
\det \tilde{g}_{ab} \ , \quad  F\equiv
F_{\tau\sigma} \ ,
 \nonumber \\
\end{eqnarray}
where $\tilde{g}_{ab}$ is the metric on
$S^2$ sphere. Let us now study the
equations of motion for given ansatz.
Firstly, the equation of motion for
$A_\alpha$ implies
\begin{equation}
\frac{(2\pi\alpha')F\sin^2\alpha}{\sqrt{\frac{R^4}{\sigma^4}
-(2\pi\alpha')^2F^2}}=\Pi
\end{equation}
and consequently
\begin{eqnarray}
2\pi\alpha'F=-\frac{R^2\Pi}
{\sigma^2\sqrt{\sin^4\alpha+ \Pi^2}}
\ , \nonumber \\
\sqrt{\frac{R^4}{\sigma^4}-(2\pi\alpha')^2F^2}=
\frac{R^2\sin^2\alpha}{\sigma^2
\sqrt{\sin^4\alpha+\Pi^2}} \ .
\nonumber \\
\end{eqnarray}
 Then in the similar way as in D5-brane
 case it is possible to show that the
equation of motion for $x^-$ implies
\begin{equation}
\partial_\tau^2
u-\frac{2}{\sigma}\partial_\sigma u+
\partial_\sigma^2 u=0 \
\end{equation}
that has again the same form as
equation given in
\cite{Ishizeki:2008dw}. Further, using
the linearity of this equation it is
possible for given profile of Wilson
loop on the boundary specified by curve
$u(0,\tau)$ find corresponding function
$u(\sigma,\tau)$.

As the next step we consider remaining
equations of motion. In fact  the
equation of motion for $\alpha$ has two
solutions, one where $\alpha_0=0$ that
corresponds to D3-brane collapsed to a
point and the second one
$\alpha_0=\frac{\pi}{2}$ corresponding
D3-brane wrapped $S^2$. In fact this is
solution we are interested in. Finally
it is easy to see that remaining
equations of motion are satisfied.

Finally we evaluate the action on
given solution and we obtain
\begin{equation}
S_{bulk}=\frac{2N}{\pi}\int d\tau
d\sigma \frac{\sin^4\alpha_0}
{\sigma^2\sqrt{\sin^4\alpha_0+\Pi^2}} \
.
\end{equation}
Again as in  case of D5-brane we have
to include boundary terms to achieve
gauge invariance. To do this
 we include to the action the
boundary expression
\begin{eqnarray}
 S_{gauge}&=& -\tau_3\int d\tau
d\sigma d\mathrm{Vol}S^2 \Pi
(2\pi\alpha')F=\nonumber \\
&=&-\frac{2N}{\pi}\int d\tau d\sigma
\frac{\Pi^2}{\sigma^2
\sqrt{\sin^4\alpha_0+\Pi^2}} \ .
\nonumber \\
\end{eqnarray}
Then the on-shell action is equal to
\begin{eqnarray}
S&=&S_{bulk}+S_{gauge}=
\frac{2N}{\pi}\int d\tau d\sigma
\frac{1}{\sigma^2}
\sqrt{\sin^4\alpha_0+\Pi^2}=\nonumber \\
&=&
\frac{2N}{\pi}\sqrt{\sin^4\alpha_0+\Pi^2}
\int d\tau d\sigma \frac{1}{\sigma^2}
=\frac{4N}{\sqrt{\lambda}}
\sqrt{1+\Pi^2} S_{funstring} \ .
\nonumber \\
\end{eqnarray}
We again see that the resulting action
is proportional to the fundamental
string action. On the other hand
 an
interpretation of this D3-brane
configuration in dual SYM is not
complete clear as was argued in
\cite{Hartnoll:2006ib}. Then we should
rather consider this solution as an
interesting example of non-trivial
D3-brane configuration that ends on the
boundary of $AdS_5$.
\section{D3-brane on
$AdS_5$}\label{third}
 In this section
we consider D3-brane that is embedded
in $AdS_5$ only. This D3-brane with
electric flux $\Pi$ should correspond
to the Wilson line in rank $\Pi$
symmetric representation.
As opposite to the examples studied in
previous section we start with the
D3-brane action where the world-volume
has Minkowski signature so that the
action takes the form
\begin{equation}
S_{D3}= -\tau_3 \int d^4\xi
\sqrt{-\det\bA}+\frac{ \tau_3}{4!} \int
C_{MNPQ} dX^M\wedge dX^N \wedge dX^P
\wedge dX^Q \ ,
\end{equation}
where
\begin{equation}
C^{(4)}=
\frac{R^2}{z^4} dx_+ \wedge dx_-\wedge
dx^1 \wedge dx^2 \ .
\end{equation}
 To begin with we
 introduce light-cone variables $\xi^\pm=
\frac{1}{\sqrt{2}}(\xi^1\pm \xi^0)$
and  consider following ansatz
\begin{eqnarray}\label{ansD3}
z= z(\xi^3)\equiv z(\sigma) \ ,  \quad
X^1=\xi^2\equiv \tau , \quad X^+=\xi^+
\ , \quad  X^-=\xi^- \ , \quad
X^3=\xi^3
\nonumber \\
\end{eqnarray}
with non-trivial gauge field strengths
$F_{\tau\sigma}=-F_{\sigma\tau}=F
 \ , F_{+\sigma}=-F_{\sigma +}=\mF$.
For this ansatz the matrix
$\bA_{\alpha\beta}$ takes the form
\begin{eqnarray}\label{bA}
\bA_{++}&=&0
\ , \quad
\bA_{--}=0 \ , \nonumber \\
\bA_{+-}&=&\frac{R^2}{z^2} \ , \quad
\bA_{-+}=\frac{R^2}{z^2} \ ,
\nonumber \\
\bA_{23}&=&-\bA_{32}=
(2\pi\alpha')F \ , \nonumber \\
\bA_{+3}&=&-\bA_{3+}=(2\pi\alpha')\mF \
,
\nonumber \\
\bA_{22}&=&\frac{R^2}{z^2} \ , \quad
\bA_{33}=\frac{R^2}{z^2}(1+z'^2) \ ,  \nonumber \\
\end{eqnarray}
where $z'=\partial_\sigma z$.  Then we
easily obtain
\begin{equation}
\det\bA=-\frac{R^8}{z^8}(1+z'^2)-\frac{R^4}{z^4}
(2\pi\alpha')^2F^2 \ .
\end{equation}
 In order to solve the equations of
motion we have to find components of
the inverse matrix $\bAi$. Using
(\ref{bA}) we obtain
\begin{eqnarray}\label{bAiD3}
\bAi^{++}&=& 0 \ , \quad
\bAi^{--}=\frac{R^2(2\pi\alpha'\mF)^2}
{z^2(\frac{R^8}{z^8}(1+z'^2)+\frac{R^4}{z^4}
(2\pi\alpha')^2F^2)} \ , \nonumber \\
\bAi^{+-}&=&\bAi^{-+}=\frac{R^6(1+z'^2)}
{z^6[\frac{R^8}{z^8}(1+z'^2)+\frac{R^4}{z^4}
(2\pi\alpha')^2F^2]} \ , \nonumber \\
\bAi^{2+}&=&\bAi^{+2}=0=
\bAi^{+3}=\bAi^{3+}=0 \ ,
  \nonumber \\
\bAi^{-2}&=&\bAi^{2-}=-\frac{R^2(2\pi\alpha')^2\mF
F}
{z^2[\frac{R^8}{z^8}(1+z'^2)+\frac{R^4}{z^4}
(2\pi\alpha')^2F^2]} \ ,  \nonumber \\
\bAi^{-3}&=&\bAi^{3-}=-\frac{R^4(2\pi\alpha')\mF}
{z^4[\frac{R^8}{z^8}(1+z'^2)+\frac{R^4}{z^4}
(2\pi\alpha')^2F^2]} \ , \nonumber \\
\bAi^{22}&=&-\frac{R^6(1+z'^2)}{z^6
[\frac{R^8}{z^8}(1+z'^2)+\frac{R^4}{z^4}
(2\pi\alpha')^2F^2]} \ , \nonumber \\
\bAi^{33}&=&\frac{R^6}{z^6
 [\frac{R^8}{z^8}(1+z'^2)+\frac{R^4}{z^4}
(2\pi\alpha')^2F^2]} \ , \nonumber \\
\bAi^{23}&=&-\bAi^{32}=-\frac{R^4(2\pi\alpha')F}{
z^4[\frac{R^8}{z^8}(1+z'^2)+\frac{R^4}{z^4}
(2\pi\alpha')^2F^2]} \ .
\nonumber \\
\end{eqnarray}
Then using (\ref{bAiD3}) it is easy to
see that the equations of motion for
$A_+, A_-$ are obeyed automatically. On
the other hand the equation of motion
for $A_2,A_3$ implies an existence of
conserved electric flux $\Pi$
\begin{eqnarray}
\frac{ (2\pi\alpha')F}
{\sqrt{(1+z'^2)+\frac{z^4}{R^4}
(2\pi\alpha')^2F^2}}=\Pi \nonumber \\
\end{eqnarray}
and consequently
\begin{eqnarray}
(2\pi\alpha'F)^2&=&
\frac{\Pi^2(1+z'^2)}
{1-\frac{z^4}{R^4}\Pi^2} \ ,
\nonumber \\
\sqrt{1+z'^2+\frac{z^4}{R^4}(2\pi\alpha'F)^2}&=&
\sqrt{\frac{1+z'^2}{1-\frac{z^4}{R^4}\Pi^2}} \ .  \nonumber \\
\end{eqnarray}
Using these relations we obtain that
the equation of motion for $X^+$ takes
the form
\begin{eqnarray}\label{eqX+}
& &\partial_3\left[\frac{R^2
(2\pi\alpha') \mF}{z^2
\sqrt{(1+z'^2)+\frac{z^4}{R^4}(2\pi\alpha'F)^2}}\right]+
\partial_2\left[\frac{ (2\pi\alpha')^2 F\mF}
{\sqrt{(1+z'^2)+\frac{z^4}{R^4}(2\pi\alpha'F)^2}}\right]=
\nonumber \\
&=&\partial_\sigma\left[\frac{R^2\sqrt{1-\frac{z^4\Pi^2}{R^4}}
\mF}{z^2
\sqrt{1+z'^2}}\right]+\Pi\partial_\tau
\mF=0 \ .
\nonumber \\
\end{eqnarray}
We return to implications of this
equation below. As the next step we
consider  the equation of motion for
$X^3$ and we obtain
\begin{eqnarray}
& &
\partial_3[g_{33}\bAi^{33}_S\sqrt{-\det\bA}]
-\partial_3[C_{+-23}]=\nonumber \\
&=&\partial_\sigma
\left[\frac{R^4}{z^4}\left(
\frac{\sqrt{1-\frac{z^4}{R^4}\Pi^2}}
{\sqrt{1+z'^2}}-1\right)\right]=0
\nonumber \\
\end{eqnarray}
and consequently
\begin{equation}\label{difz}
\frac{\sqrt{1-\frac{z^4}{R^4}\Pi^2}}
{\sqrt{1+z'^2}}=1+\frac{z^4}{R^4}B \ ,
\nonumber \\
\end{equation}
where $B$ is an integration constant.
In what follows we consider the case
when
 $B=0$ and introduce complex electric
 flux as $\Pi=i\tilde{\Pi}$. Then
 the equation (\ref{difz}) takes simple
 form
 \begin{equation}\label{z'sim}
z'=-\frac{z^2}{R^2}\tilde{\Pi}
\end{equation}
that can be easily integrated with the
result
\begin{equation}\label{zsigma}
\frac{1}{z}=\frac{\tilde{\Pi}}{R^2}\sigma
\end{equation}
with appropriate chosen integration
constant.

Now we return to the equation
(\ref{eqX+}). Using (\ref{z'sim}) and
(\ref{zsigma}) this equation simplifies
considerably
\begin{equation}
\frac{\tilde{\Pi}^2}{R^2}\partial_\sigma (\sigma^2\mF)+
i\tilde{\Pi}\partial_\tau \mF=0
\end{equation}
Since  $\mF_{+\sigma}=\mF=-\partial_\sigma u_+$ the equation
above implies
\begin{eqnarray}\label{equd3}
\frac{\tilde{\Pi}}{R^2}
\sigma^2\partial_\sigma u_++
i \partial_\tau u_+=0
\nonumber \\
\end{eqnarray}
and we solve this equation with the
ansatz
\begin{eqnarray}\label{anstu}
u_+(\sigma,\tau)=\int_0^{\infty}
d\omega[e^{i\omega\tau}
u_+(\sigma,\omega)+e^{-i\omega\tau}u^*_+(\sigma,\omega)]
\end{eqnarray}
so that $u^*_+=u_+$. If we insert
(\ref{anstu}) into (\ref{equd3}) we
obtain
\begin{equation}
\frac{\tilde{\Pi}}{R^2}\sigma^2 u'_+-
\omega u_+=0
\end{equation}
with the solution
\begin{equation}
u_+=C_\omega e^{-\omega
\frac{R^2}{\sigma^2\tilde{\Pi}}}=
C_\omega e^{-\omega z} \ .
\end{equation}
Then the general solution (\ref{anstu})
takes the form
\begin{equation}
u_+(z,\tau)=\int_0^\infty d\omega
e^{-\omega z} [e^{i\omega
\tau}C_\omega+e^{-i\omega
\tau}C^*_\omega] \ ,
\end{equation}
where we can determine $C_\omega$ using
the known profile $u_+(z=0,\tau)$ of
Wilson line on the boundary $z=0$.
 As  final remark
note that can be easily shown that the
equations of motion for $X^2$ and $Z$
are obeyed for the ansatz
(\ref{ansD3}).

Let us finally
evaluate the action on solution found
above.
If we calculate
 the action for this
solution we find that the WZ term
exactly cancels the DBI part and
consequently $S=0$. Of course this is
expected  final answer but we should
take the boundary terms into account as
well. Careful analysis of this problem
was presented in \cite{Drukker:2005kx}
with the result that the boundary terms
for this particular D3-brane
configurations do not contribute to the
action.

Let us outline our results. We found
solutions of D3-brane equations of
motion that is an analogue of the
fundamental string solution presented
in \cite{Ishizeki:2008dw} and that
should be related to the expectation
values of Wilson loops evaluated in
$\Pi$-symmetric representation in dual
$N=4$ SYM theory.
 We again
mean that this fact nicely shows an
efficiently of the D3-brane description
of configurations with large number of
fundamental strings.
\\
\\
{\bf Acknowledgement}

This work
 was supported  by the Czech Ministry of
Education under Contract No. MSM
0021622409.

\begin{appendix}\label{appendix}
\section{Dp-brane in general background}
In this appendix we review the form of
  Dp-brane action in general background. We
also review corresponding equations of
motion.

As is well known the Dp-brane action in
general background consists two parts,
Dirac-Born-Infeld part (DBI) and
Wess-Zumino part (WZ) so that
\begin{eqnarray}\label{actD1}
S&=&S_{DBI}+S_{WZ} \ , \nonumber \\
S_{DBI}&=&-\tau_p
\int d^{p+1}\xi e^{-\Phi}
\sqrt{-\det\bA} \ , \nonumber \\
\bA_{\alpha\beta}&=&\partial_\alpha
X^M\partial_\beta X^N
G_{MN}+(2\pi\alpha')\mF_{\alpha\beta}  \ , \nonumber \\
  \mF_{\alpha\beta}&=&
\partial_\alpha A_\beta-\partial_\beta A_\alpha-
(2\pi\alpha')^{-1}B_{MN}\partial_\alpha
X^M\partial_\beta X^N
 \ , \nonumber \\
S_{WZ}&=&\tau_p\int e^{(2\pi\alpha')\mF}\wedge C=
\tau_p\sum_{n\leq 0}
\frac{(2\pi\alpha')^n}{n! (2!)^n q!}
\int d^{p+1}\xi
e^{i_1\dots i_{p+1}}
(\mF^n)_{i_1\dots i_{2n}}
C_{i_{2n+1}\dots i_{p+1}}
 \ ,  \nonumber \\
\end{eqnarray}
where $\tau_p$ is Dp-brane tension,
$\xi^\alpha,\alpha=0,\dots,p$ are
world-volume coordinates and where
$A_\alpha$ is gauge field living on the
world-volume of Dp-brane. Note also
that $C$ in the last line in
(\ref{actD1}) means collection of
Ramond-Ramond forms where $q=p+1-2n$.

If we now perform the variation of
(\ref{actD1}) with respect to $X^M$ we
obtain following equations of motion
for  $X^M$
\begin{eqnarray}\label{eqxm}
& &-\tau_p\partial_M[ e^{-\Phi}]
\sqrt{-\det\bA}\nonumber \\
&-&\frac{\tau_p}{2}
e^{-\Phi}(\partial_M
g_{KL}\partial_\alpha X^K
\partial_\beta X^L-
b_{KL}
\partial_\alpha X^K\partial_\beta X^L)\bAi^{\beta\alpha}
\sqrt{-\det\bA}+\nonumber \\
&+&\tau_p\partial_\alpha
[e^{-\Phi}g_{MN}\partial_\beta
X^N\bAi_S^{\beta\alpha}\sqrt{-\det\bA}]-
\nonumber \\
&-&\tau_p
 \partial_\alpha
[e^{-\Phi}b_{MN}\partial_\beta X^N
\bAi_A^{\beta\alpha}\sqrt{-\det\bA}]+J_M=0 \ , \nonumber \\
\end{eqnarray}
where
\begin{eqnarray}
J_M=\frac{\delta S_{WZ}}{\delta X^M}
=\tau_p\sum_{n\leq 0}
\frac{(2\pi\alpha')^n}{n! (2!)^n q!}
\epsilon^{i_1\dots i_{p+1}}
(\mF)^n_{i_1 \dots i_{2n}}
\partial_{i_{2n+1}}X^{M_{2n+1}}
\dots \partial_{i_{p+1}}X^{M_{p+1}}
F_{M M_{2n+1}\dots M_{p+1}}
 \ . \nonumber \\
\end{eqnarray}
In the same way the variation of (\ref{actD1})
with respect to  $A_\alpha$ implies following
equation of motion
\begin{equation}\label{eqa}
2\pi\alpha'\tau_p
\partial_\alpha [e^{-\Phi}\bAi^{\beta\alpha}_A\sqrt{-\det\bA}]+
J_\alpha=0 \ ,
\end{equation}
where
\begin{eqnarray}
J_\alpha &=&\frac{\delta S_{WZ}}{\delta
A_\alpha}= \nonumber \\
&=&\epsilon^{i_1\dots i_{p+1}}
\sum_{n\geq 0}
\frac{(2\pi\alpha')^n}{n! (2!)^n
(q-1)!}(\mF)^n_{i_2\dots i_{2n+1}}
\partial_{i_{2n+2}}X^{M_{2n+2}}
\dots \partial_{i_{p+1}}X^{M_{p+1}}
F_{M_{2n+2}\dots M_{p+1}} \ ,  \nonumber \\
\end{eqnarray}
and where we have also defined
\begin{equation}
\bAi_S^{\alpha\beta}=
\frac{1}{2}\left(\bAi^{\alpha\beta}+\bAi^{\beta\alpha}\right) \ ,
\quad
\bAi_A^{\alpha\beta}=
\frac{1}{2}\left(\bAi^{\alpha\beta}-\bAi^{\beta\alpha}\right) \ .
\end{equation}

\end{appendix}


\end{document}